\documentclass{ws-procs9x6}

\def\numunue{\nu_\mu\rightarrow\nu_e}
\def\numunutau{\nu_\mu\rightarrow\nu_\tau}

\def\nutau{\nu_\tau}

\def\numu{\nu_\mu}

\def\pot{\textit{pot}}

\begin{document}

\title{The OPERA experiment: \\ a direct search of the $\nu_\mu \longrightarrow
\nu_\tau$ oscillations}

\author{J.Marteau$^*$  for the OPERA collaboration}

\address{Institut de Physique Nucléaire de Lyon, UCBL-CNRS/IN2P3,\\ 
4 rue E.Fermi, F-69622 Villeurbanne, France\\
$^*$E-mail: marteau@ipnl.in2p3.fr}

\begin{abstract}
The aim of the OPERA experiment is to search for the appearance of the tau neutrino in the quasi pure muon neutrino beam produced at CERN (CNGS). The detector, installed in the Gran Sasso underground laboratory 730~km away from CERN, consists of a lead/emulsion target complemented with electronic detectors. 
After a short pilot run in 2007, a first physics run took place from June to November 2008. The second physics run started in June 2009. At present a total (2008+2009) of ~4.2 10$^{19}$ protons on target were delivered by the CNGS, producing more than 25,000 events in time coincidence in the OPERA detector. Among them \~4000 events occured in the target of the detector. In this paper the detector and the analysis strategy are described and the status of the analysis of the 2008 and 2009 runs is discussed.
\end{abstract}


\bodymatter

\section{Introduction}\label{sec:intro}
In the last decades solar and atmospheric neutrino experiments observed deficits in the measured fluxes which are all
well reproduced in a neutrino oscillations model, implying non vanishing, not degenerate neutrino masses and neutrino mixing.
Within such hypothesis weak interactions eigenstates differ from the mass eigenstates. The mixing can be parametrized in an unitary
matrix whose parameters (3 angles and 1 or 3 phases depending on the Dirac or Majorana nature of neutrinos) associated to the square
masses differences $\Delta m^2$ drive the amplitude of the disappearance ($P(\nu_\alpha \rightarrow \nu_\beta)$) or survival 
($P(\nu_\alpha \rightarrow \nu_\alpha)$) probabilities. 
 
The major experimental results for solar\cite{chlore,gno,sage,sk,kam,imb,sno}, atmospheric\cite{sk,kam,imb,sno,soudan2,macro}, reactor~\cite{chooz,kamland} or accelerator~\cite{k2k,minos} neutrinos were obtained by observing the ``disappearance" of neutrinos wrt to a close position measurement or a predicted flux. OPERA~\cite{opera} has been designed to perform an unique \underline{appearance} observation of the oscillation products to confirm (or infirm) the neutrino oscillation hypothesis in the atmospheric sector through the $\numunutau$ channel and also to set limits on the $\theta_{13}$ angle through the $\numunue$ channel. 

\section{The OPERA experiment}
\subsection{The CNGS beam}\label{subsec:cngs}
\begin{figure}[!hb]
\begin{center}
\psfig{file=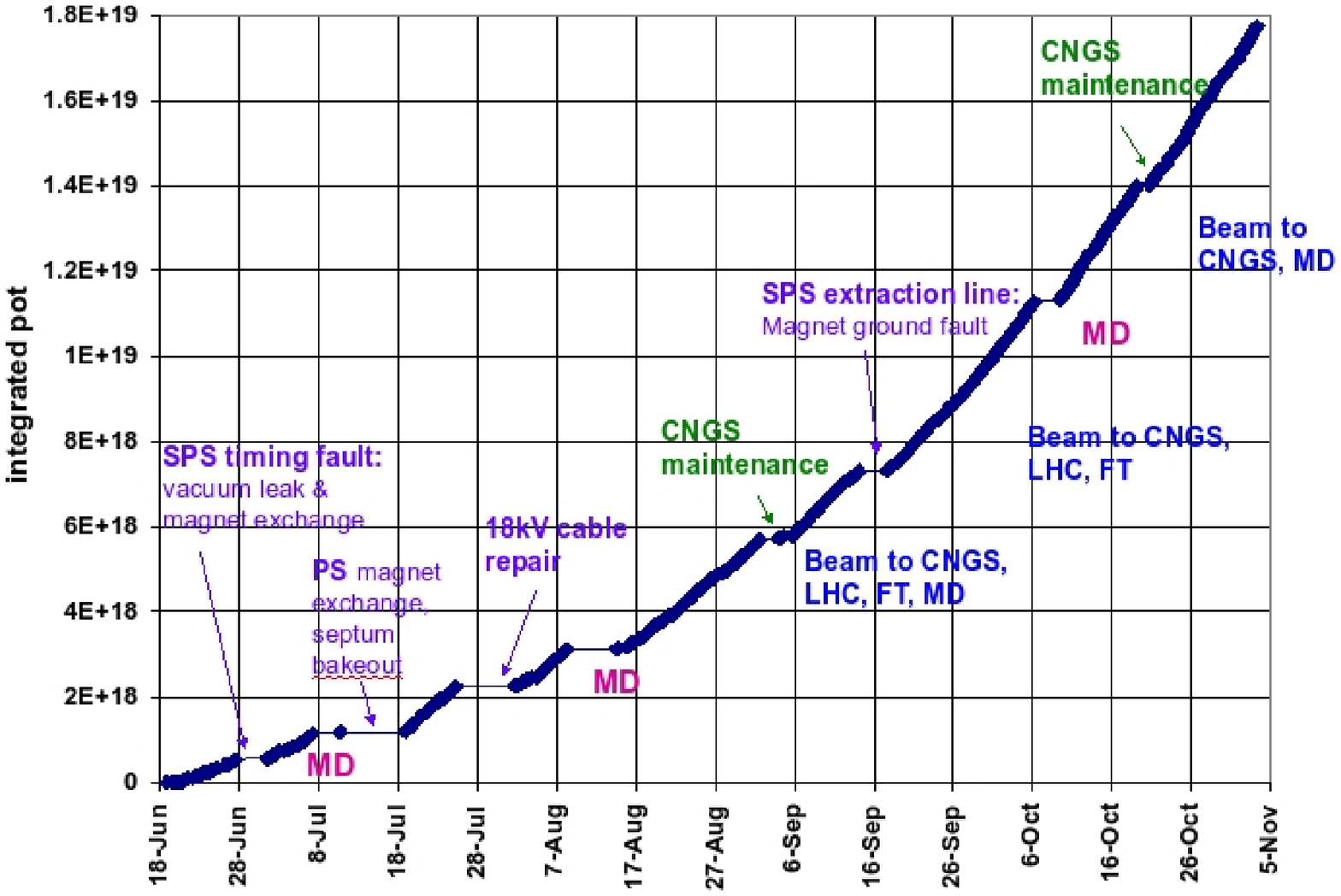,height=5.5cm,width=5.5cm} \hfill
\psfig{file=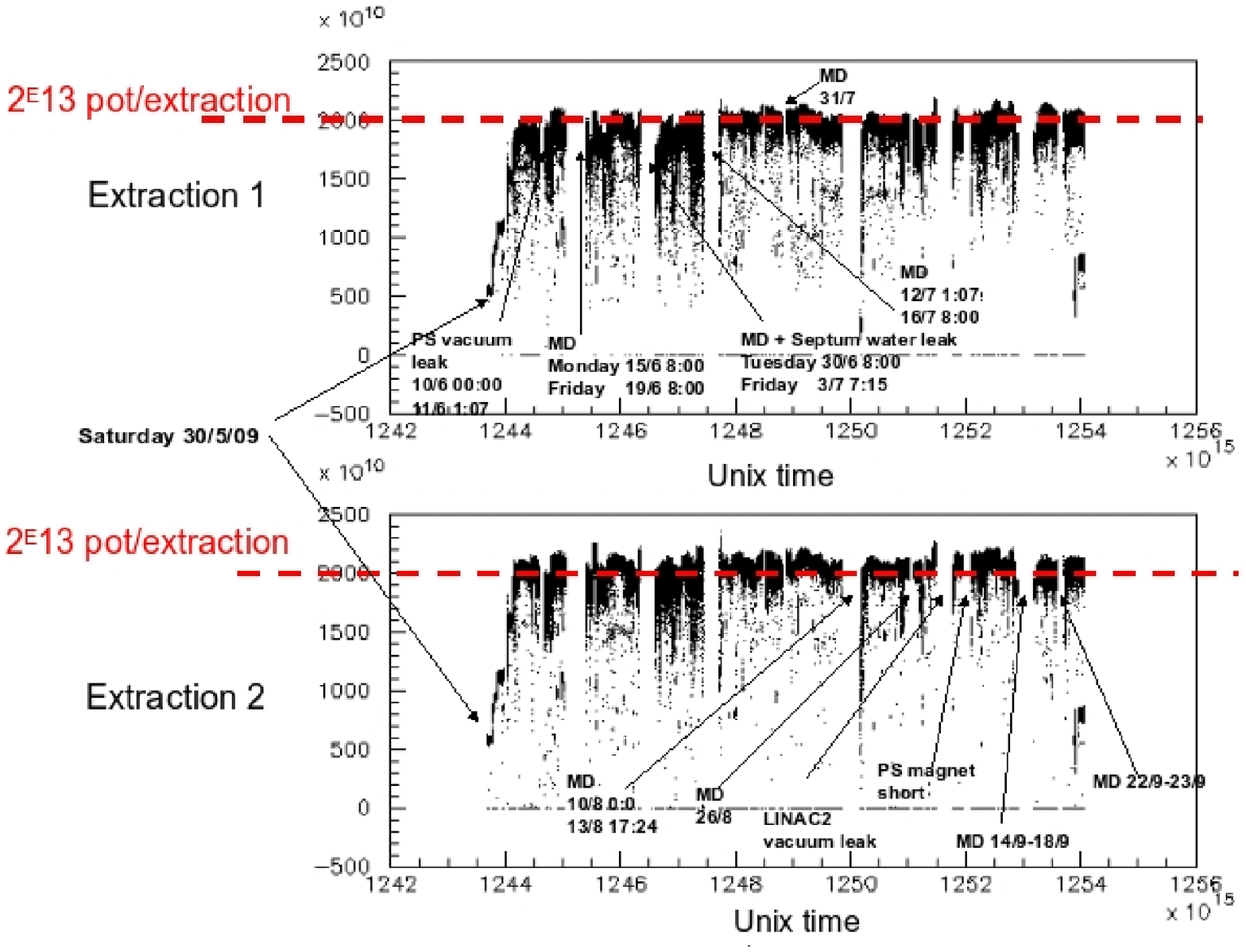,height=5.5cm,width=5.5cm}
\end{center}
\caption{Integrated \pot delivery versus time during 2008 run (left). Protons per extraction during 2009 run (right).}
\label{fig:cngs}
\end{figure}

The CNGS~\cite{cngs} programme of neutrino beam from CERN to Gran Sasso has been approved in 1999.  
The beam has been optimized to maximize the number of $\tau$ events in the detector (convolution of the neutrino flux, the appearance probability and the detection efficiency). The neutrino average energy is 17~GeV. Measured in the number of interactions in the detector, the $\bar{\nu_\mu}$ contamination is $\sim 2\%$, the $\nu_e$ ($\bar{\nu_e}$) is $< 1\%$ and the number of prompt $\nu_\tau$ is negligible. 
The beam configuration starts from the SPS protons directed to the target chamber where mainly pions and kaons are produced, which decay then into a ~1km long decay tunnel followed by hadron stops and muon detectors. From CERN to Gran Sasso the escaping neutrinos travel for 2.44~ms and their mean direction w.r.t. the horizontal in Gran Sasso forms a 3$^\circ$ angle due to the earth curvature. 
There are two fast extractions separated by 50~ms in each CNGS cycle (6s long). These cycles are repeated 3 or 4 times during each SPS supercycle. The nomimal beam intensity is $4.5 \cdot 10^{19}$ \pot/year. The so far achieved performance in 2008 and 2009 is respectively 1.78$^{19}$  and 2.4$^{19}$ $\pot$ (Fig.~\ref{fig:cngs}).  

\subsection{The detection technique}\label{subsec:ecc}
The challenge of the experiment is to measure the appearance of $\nutau$ from $\numu$ oscillations through CC $\tau$ interactions. The events induced by the short-lived $\tau$ have a characteristic topology (with a ``kink'' due to the presence of undetected neutrinos in the $\tau$ decay) but extends over $\sim mm^3$ typical volumes. 
The detector should therefore match a large mass for statistics, a high spatial resolution and high rejection power to limit background contamination. These requirements are satisfied using the proven ECC (Emulsion Cloud Chamber) technique which already worked successfully in the DONUT experiment~\cite{donut}. The passive target consists of lead plates. Particles are tracked in nuclear emulsions films with a sub-micrometric intrinsic resolution. 57 emulsions films are assembled and interspaced with 56 lead plates 1~mm wide in a detector basic cell called ``brick''. An additional doublet of emulsion film (Changeable Sheets, CS) is attached on the downstream face of each brick to guide the tracks predictions inside the brick itself. A brick is a 12.7 $\times$ 10.2 cm$^2$ object with a thickness along the beam direction of 7.5 cm (about 10 radiation lengths). 
Bricks are assembled in 31 walls ($52 \times 64$ bricks) separated by electronic detectors planes to trigger the event and identify the brick with the interaction vertex. The total OPERA target contains ~150000 bricks with a total mass of 1.25 ktons. The ECC bricks were assembled
underground at an average rate of 700 per day by a dedicated fully automated Brick Assembly Machine (BAM); the OPERA target was filled using two automated manipulator systems (BMS). The passage of a m.i.p. in an emulsion film results, after development, in a set of aligned grains, ~35 grains /100$\mu$m. 

\subsection{The off-line emulsions scanning}
After development emulsions are scanned by automatic microscopes whose nominal speed is higher than $\sim 20$ cm$^2$/h per emulsion layer ($44 ~\mu$m thick). There are two different approaches developed by the OPERA collaboration, in Europe (ESS~\cite{ESS}, based on software image reconstruction) and in Japan (S-UTS~\cite{SUTS} based on hard-coded algorithms). 
The scanning sequence proceeds with the division of the emulsion thickness into $\sim$~16 tomographic images by focal plane adjustment, images digitization and track finding algorithms. Track grains are identified and separated from ``fog'' grains and associated into ``micro-tracks''. 
\begin{figure}[!hb]
\begin{center}
\psfig{figure=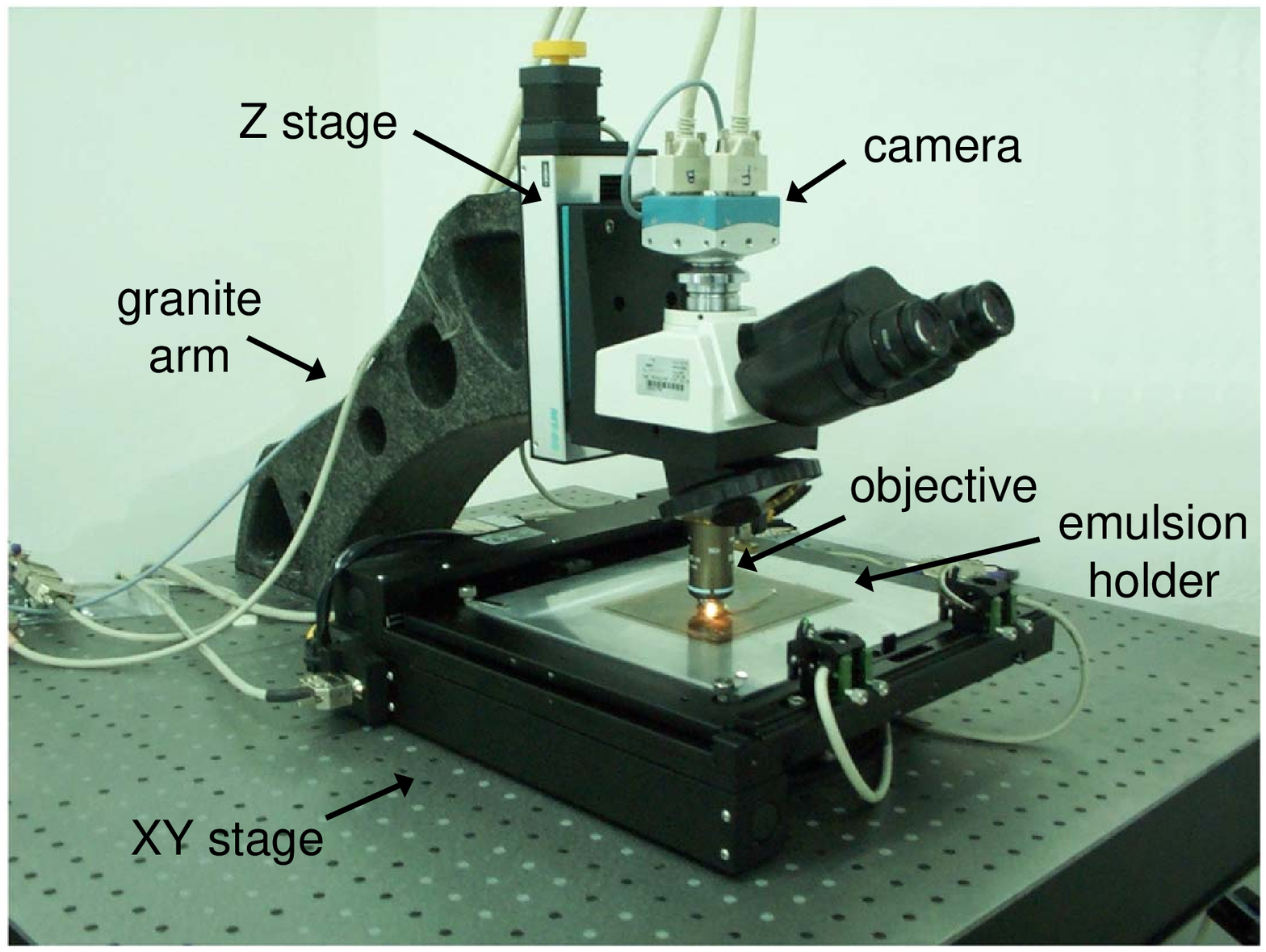,height=4cm,width=5.5cm} \hfill
\psfig{figure=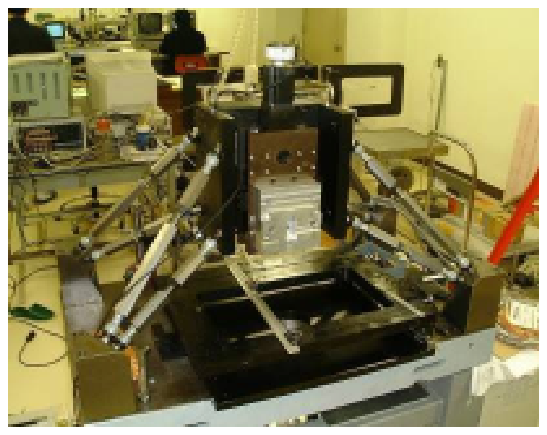,height=4cm,width=5.5cm}
\end{center}
\caption{Pictures of one of the ESS microscopes (left) and of the S-UTS (right).}  
\label{fig:scanning}
\end{figure}

\subsection{The event validation chain}
Every time a trigger in the electronic detectors (see later) is compatible with an interaction inside the target and on time with the beam arrival, the brick with the highest probability to contain the neutrino interaction vertex is extracted from the apparatus and exposed to X-rays for film-to-film alignment between the CS and the last brick film. CS films are scanned looking for tracks in an area of about 50 cm$^2$ around the TT prediction. The CS-to-TT connection is made with an accuracy of ~1 cm in position and 20 mrad in angle. If tracks from the CS scanning are compatible with those reconstructed in the electronic detectors the brick is brought to the surface laboratory and exposed to cosmic rays in a dedicated pit in order to provide straight tracks for a finer (sub-micrometric) film-to-film alignment. Brick emulsion films are then developed and dispatched to the scanning laboratories to be analysed. 
All tracks reconstructed in the CS are sought in the most downstream films of the brick with an accuracy of about 70 $\mu$m in position and 10 mrad in angle. Confirmed tracks are followed back along the brick. A stopping point is a signature either of a primary or a secondary vertex. Vertex confirmation is made by scanning a volume of 1 cm$^2$ transverse size for 11 films located upstream and
downstream of the stopping point. All tracks originating inside the volume are input for a vertex reconstruction algorithm which is tuned to find also decay topologies. The tracks attached to the primary vertex are followed downstream to check whether they show any decay topology and to measure their momentum from their Coulomb multiple scattering.

\subsection{Expected performances} 
At the expected nominal beam intensity and for five years data taking a total of ~25000 charged and neutral current interactions is expected in the nominal mass target of OPERA. Among these about ~100 $\nutau$ CC interactions are expected for oscillation parameter values $\Delta m^2_{23}$=2.4~$\times$~10$^{-3}$~eV$^2$ and sin$^2$2$\theta_{23}$=1. The overall detection efficiencies have been estimated in Monte Carlo simulations and are currently under re-evaluation by exploiting the data themselves. A total of 10 identified $\tau$ events is expected (in all physics channels) for less than 1 background event.
%
In the sub-dominant $\numunue$ channel OPERA should, in the same run conditions, set a limit of $\sin^2 2\theta_{13}<0.06$ (90\%~C.L.) \cite{t13bib}, the main militation being the beam contamination in $^{-}\nu_e$.   
 
\subsection{The OPERA electronic detectors}\label{subsec:detector}
\begin{figure}[!ht]
\begin{center}
\psfig{figure=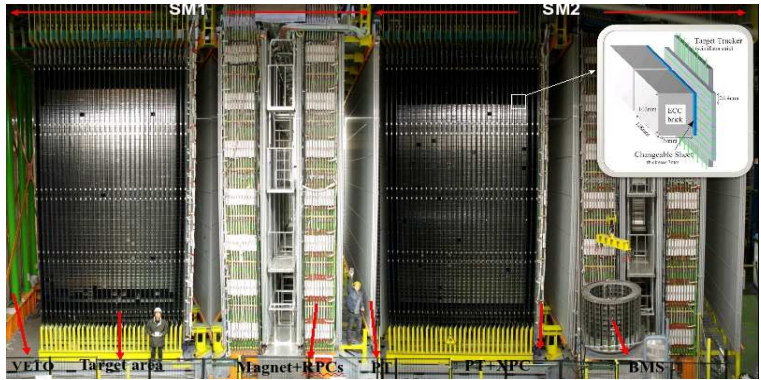,height=5cm,width=10cm}
\end{center}
\caption{Side view of the OPERA detector. The two target modules, 31 bricks walls spaced by black 
covered scintillator biplanes, are separated by the first spectrometer, 6 stations of drift tubes upstream, between
and downstream the two arms of a dipolar magnet instrumented with RPC.}  
\label{fig:opera}
\end{figure}

OPERA is a hybrid detector made of two identical Super Modules (SM1 and SM2) consisting of a target section followed by a muon spectrometer (see Fig.~\ref{fig:opera}). The target section is made of the already mentionned 31 brick walls each one being followed by a highly segmented scintillator tracker plane. A large VETO plane is placed in front of the detector to further discriminate beam events from horizontal cosmics and beam neutrino interactions in the rock. The construction of the experiment started in Spring 2003. The two instrumented magnets were completed in May 2004 and beginning of 2005 respectively. In Spring 2006 all scintillator planes were installed.  

The target tracker covers a total area of 7000~m$^2$ and is built of 32000 scintillator strips, each 7~m long and of 25~mm$ \times $15~mm cross section. Along the strip, a wavelength shifting fiber of 1~mm diameter transmits the light signals to both ends, where the opto-electonics readout chain includes the Hamamatsu 64 channels MaPMT, a dedicated auto-triggerable front-end ASIC~\cite{asic} allowing gain correction and an Ethernet capable microprocessor digital board~\cite{elba}. The muon spectrometer consists of a large $8 \times 8$~m$^2$ dipolar magnet delivering a magnetic field of 1.55~T and instrumented with RPC's and drift tubes. Each magnet arm consists of twelve 5~cm thick iron slabs, alternating with RPC planes. This sandwich structure allows the tracking in the magnetic field to identify the muons and to determine their momentum and sign. In addition the precision tracker \cite{specbib} measures the muon track coordinates in the horizontal plane. It is made of 8~m long drift tubes with an outer diameter of 38~mm. The charge misidentification is expected to be 0.1~\% - 0.3~\% in the relevant momentum range which is efficient enough to minimize the background originating from the charmed particles produced in $\nu_{\mu}$ interactions. With the muon spectrometer a momentum resolution of $\Delta p / p \le 0.25$ for all muon momenta $p$ up to a maximum of $p=25$~GeV/c can be achieved. 

\section{Status of the analysis}

This section gives an overview of the present status of the data collection and analysis in the electronic detectors and in the emulsions during the 2008 and 2009 physics run. Currently a total of 4.2 10$^{19}$ {\it pot} have been delivered by CERN, resulting in 20 (8 + 12) millions of raw events (L2 trigger), 25,000 (10,000 + 15,000) events tagged by the OPERA detector in time coincidence with the CNGS. The event timestamp in OPERA is given with a 10ns accuracy through the common clock distribution system locked on the GPS~\cite{elba}. This large statistics is used to confront MC with data results to re-evaluate the real performance of the experiment~\cite{opera1,opera2}. 

\subsection{Electronic detectors analysis}
\begin{figure}[!hb]
\begin{center}
\psfig{figure=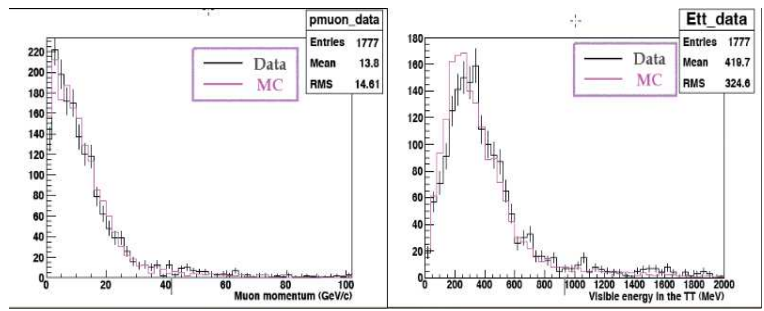,height=4cm,width=10cm} 
\end{center}
\caption{TT analysis: muon momentum reconstruction (left) and visible energy (right).}  
\label{fig:muon}
\end{figure}

Recent progresses in the reconstruction of kinematics distributions in the electronic detectors allow to improve the identification and the brick finding efficiencies (Figure~\ref{fig:muon} displays for example the good data/MC agreement for muon momentum and the visible energy in the Target Tracker). 

A lot of progress has been recorded as well in the event identification with a special emphasis on the low multiplicity events (mostly soft NC events) which intervene as background for $\tau$ physics channel with low energy deposit ($\tau \to e$ quasi-elastic for instance).   
 
\subsection{Vertex location}
The validation of events by the CS scanning is displayed on Fig.\ref{fig:scanning} (left) for the 2008 run. Also shown is the vertex distribution projected on the section of the brick (same Figure, right panel). The analysis is continuously on-going but also upgraded to improve the location efficiency. 
\begin{figure}[!ht]
\begin{center}
\psfig{figure=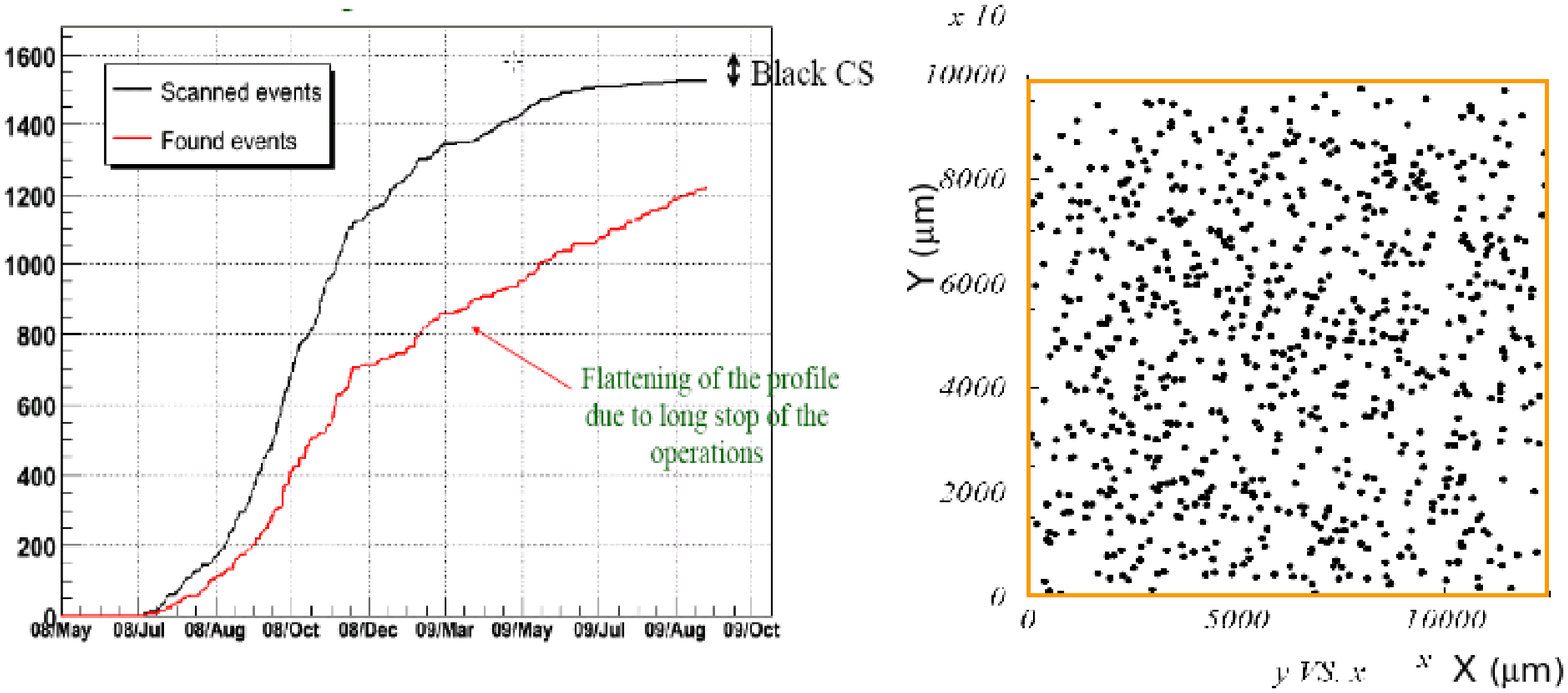,height=4.5cm,width=11cm} 
\end{center}
\caption{2008 events scanning progress: number of vertex found (left) and vertex spatial distribution (right).}  
\label{fig:scanning}
\end{figure}

\subsection{Emulsions data analysis}
\begin{figure}[!hb]
\begin{center}
\psfig{figure=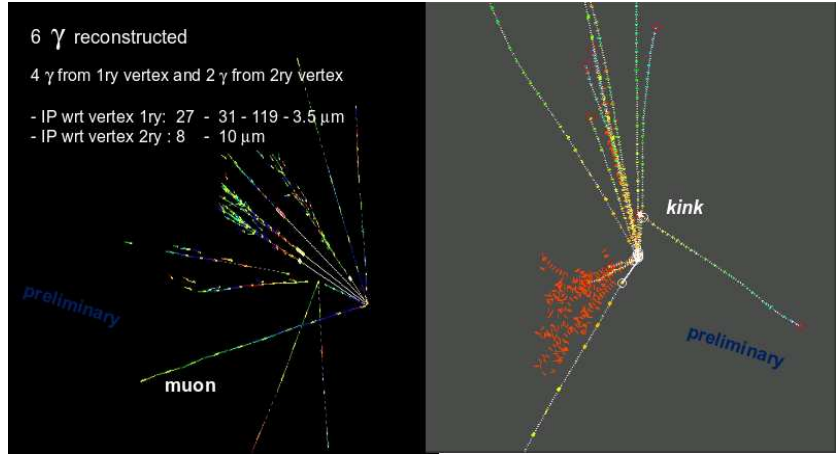,height=6cm,width=11cm} 
\end{center}
\caption{IP measurements for $\gamma$ reconstruction in the emulsions (left). Charm decay candidate (right).}  
\label{fig:scanning-evts}
\end{figure}

Important steps are under careful and systematic (re)evaluation for the validation of the data analysis. Among them the distributions of the impact parameters for the tracks wrt the primary vertex are important. Fig.~\ref{fig:scanning-evts} (left) displays an example of IP measurements for gammas attached to primary or secondary vertices.
Another crucial point is the decay search and the charm production analysis. First of all because charm decays exhibit the same topology as $\tau$ decays, therefore measuring the charm-like event reconstruction efficiency provides an important cross-check of the $\tau$ event reconstruction capability. Furthermore charm events are a potential source of background, in particular if the muon at the primary vertex is not identified. Therefore, searching for charm-decays in events with the primary muon correctly identified provides a direct measurement of this background. Charm decay topologies are searched for in the sample of located neutrino interactions (Fig.~\ref{fig:scanning-evts}, right panel). At present a set of 15 charm candidates has been identified and is deeply studied. A new systematic decay search procedure has been recently endorsed and is being applied to increase the understanding of that important data sample.

\begin{figure}[!ht]
\begin{center}
\psfig{figure=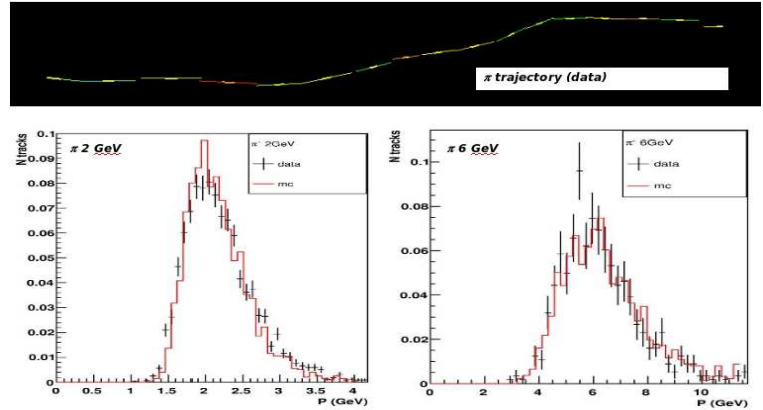,height=5.5cm,width=10cm} 
\end{center}
\caption{Pions energy reconstruction with MCS.}  
\label{fig:pions}
\end{figure}

A constant effort is brought on test beam programmes to validate the kinematics measurement by Multiple Coulomb Scattering (MCS). Tests with pion beams of different energies gave a good data/MC agreement showing the validity of the procedure (Fig.~\ref{fig:pions}) 

\subsection{Complementary studies}
On top of the oscillations physics, OPERA is also measuring the charged  ratios ($\mu^+/\mu^-$) of cosmic muons, exploiting the muon spectrometer magnetic field. An independant study of atmospheric neutrinos detection, using a time-of-flight analysis, is also performed and the up/down analysis has been shown to be efficient although the OPERA configuration is optimized for horizontal events topologies. Finally a measurement is also done in parallel of the neutrino velocity after a detailed intercalibration of the LNGS and CERN timing systems. Publications of those results are in preparation.

\section{Conclusions}
The analysis of the events collected in 2008-2009 is quite advanced. The results obtained prove that OPERA is able to accomplish the task of selecting decay topologies in the emulsions from a large number of interactions triggered by the electronic detectors and that the scene has been set for the discovery of $\nu_\tau$ appearance. The large data set recorded and under analysis allows a constant and systematic improvement of the detection procedures and efficiencies. Important data samples such as the charm decays are carefully studied since they provide crucial information in the $\tau$ appearance hunting. At the end of the 2009 run around 2 $\tau$ neutrinos events are expected...

\end{document}